\documentclass[twocolumn,showpacs]{revtex4-1}%
\usepackage{amsfonts}
\usepackage{amsmath}
\usepackage{amssymb}
\usepackage{graphicx}%
\usepackage{graphics}%
\providecommand{\U}[1]{\protect\rule{.1in}{.1in}}
\begin{document}
\title{Hidden quantum mirage by negative refraction in semiconductor P-N\ junctions}
\author{Shu-Hui Zhang$^{1}$}
\author{Jia-Ji Zhu$^{3}$}
\author{Wen Yang$^{1}$}
\email{wenyang@csrc.ac.cn}
\author{Hai-Qing Lin$^{1}$}
\author{Kai Chang$^{2,4}$}
\email{kchang@semi.ac.cn}
\affiliation{$^{1}$Beijing Computational Science Research Center, Beijing 100094, China}
\affiliation{$^{2}$SKLSM, Institute of Semiconductors, Chinese Academy of Sciences, P.O.
Box 912, Beijing 100083, China}
\affiliation{$^{3}$Institute for quantum information and spintronics, College of Science,
Chongqing University of Posts and Telecommunications, Chongqing 400065, China}
\affiliation{$^{4}$Synergetic Innovation Center of Quantum Information and Quantum Physics,
University of Science and Technology of China, Hefei, Anhui 230026, China}

\begin{abstract}
We predict a novel quantum interference based on the negative refraction
across a semiconductor P-N junction:\ with a local pump on one side of the
junction, the response of a local probe on the other side behaves as if the
disturbance emanates not from the pump but instead from its mirror image about
the junction. This phenomenon is guaranteed by translational invariance of the
system and matching of Fermi surfaces of the constituent materials, thus it is
robust against other details of the junction (e.g., junction width, potential
profile, and even disorder). The recently fabricated P-N junctions in 2D
semiconductors provide ideal platforms to explore this phenomenon and its
applications to dramatically enhance charge and spin transport as well as
carrier-mediated long-range correlation.

\end{abstract}
\maketitle

Half a century ago, Veselago proposed the concept of negative refraction for
electromagnetic waves
\cite{VeselagoSPU1968,PendryPRL2000,ZhangNatMater2008,PendryScience2012}: upon
transition from a medium with positive refractive index across a sharp
interface into a negative index medium, a diverging pencil of rays is
coherently refocused to form a sharp image or \textquotedblleft quantum
mirage\textquotedblright\ \cite{ManoharanNature2000}, similar to the bending
of light to create mirages in the atmosphere. In the past decade, negative
refraction and mirage have been observed for electromagnetic waves of various
frequencies (see Ref. \cite{ShalaevNatPhoton2007} for a review) and for cold
atoms \cite{JuzeliunasPRA2008,LederNatComm2014}. In 2007, Cheianov \textit{et
al.} \cite{CheianovScience2007} proposed the interesting idea that a sharp P-N
junction of graphene can exhibit negative refraction and hence focus electrons
out of a local pump into a sharp quantum mirage. This effect has been widely
used in theoretical proposals to control charge and/or spin transport for
massless Dirac fermions in semiconductors (see Refs.
\cite{ParkNanoLett2008,MoghaddamPRL2010,ZhaoPRL2013} for a few examples).
However, a sharp quantum mirage requires the junction to be sharp compared
with the electron wavelength ($\sim$ a few nanometers), otherwise it would
disappear due to the path-dependent phase accumulation inside the junction.
This makes the observation and application of this effect an experimentally
challenging task \cite{LeeNatPhys2015}.

In this letter, we theoretically demonstrate that in many situations where the
quantum mirage is no longer visible, its effect still exists, which could make
the response across the P-N junction independent of distance. As a basic
observation in physics, the response amplitude in a $d$-dimensional uniform
system decays at least as fast as $1/R^{(d-1)/2}$ with distance $R$,
irrespective of the energy dispersion, spin-orbit coupling, etc. This directly
leads to rapid decay of many physical properties, such as the charge and spin
conductivity \cite{IgorRMP2004} (reponse to electric/magnetic field), nonlocal
optical response \cite{CiraciScience2012}, Friedel oscillation
\cite{CheianovPRL2006,HwangPRL2008,BenaPRL2008} (response to charge impurity),
and carrier-mediated Rudermann-Kittel-Kasuya-Yosida interaction
\cite{RudermanPR1954,KasuyaPTP1956,YosidaPR1957,DietlRMP2014} (response to
magnetic impurity). The \textit{hidden} quantum mirage could lift these
constraints and dramatically enhance the nonlocal responses for electrons in
various semiconductors such as graphene, silicene, transition-metal
dichalcogenides, topological insulator surfaces, etc. As an example, we
demonstrate that the P-N junction could dramatically enhance the
carrier-mediated long-range interaction between localized magnetic moments by
several orders of magnitudes.

\begin{figure}[ptb]
\includegraphics[width=0.9\columnwidth,clip]{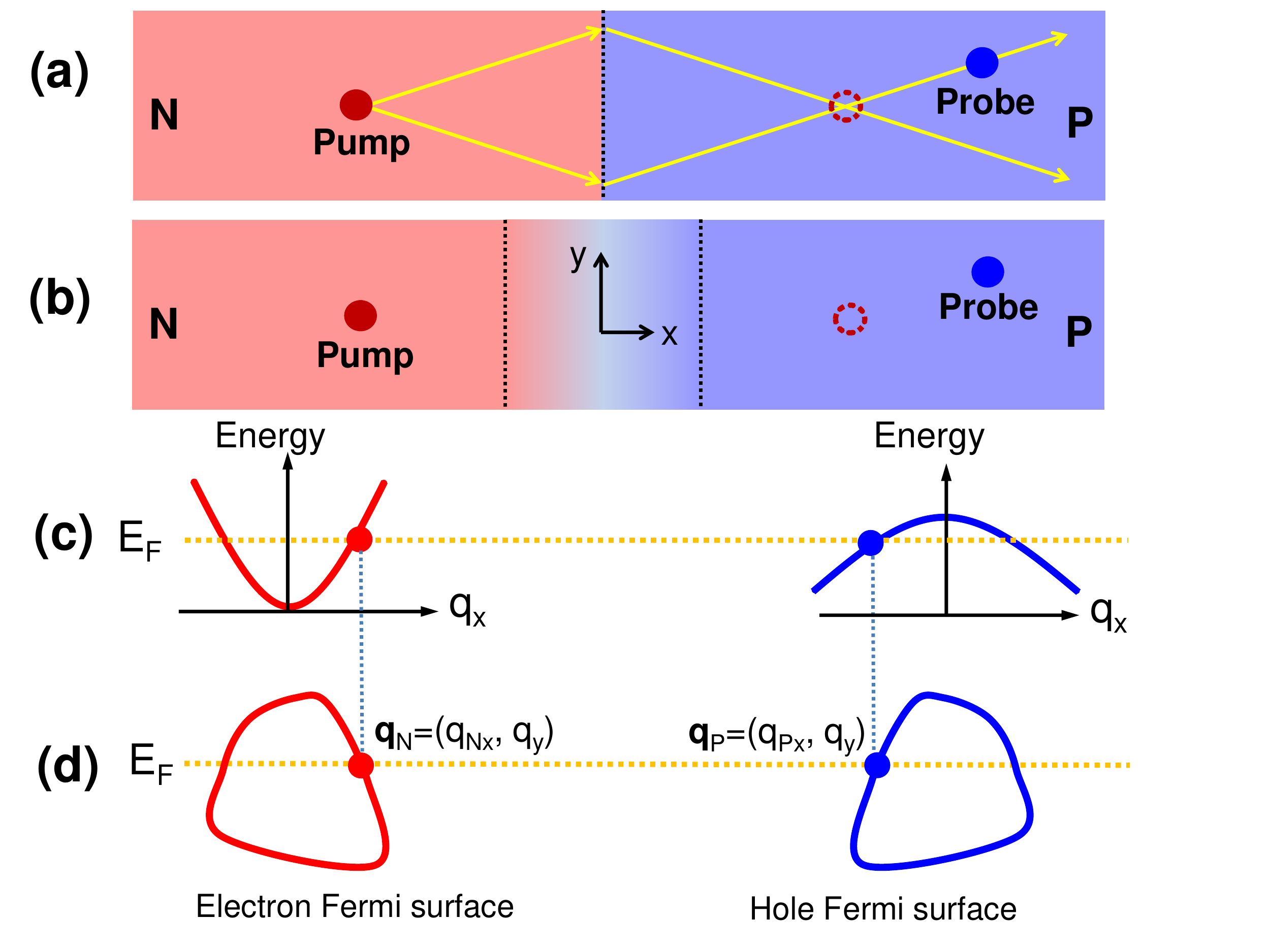}\caption{(color online).
(a)\ In a sharp P-N junction, a local pump (filled red circle) in the N region
generates a sharp quantum mirage (dashed red circle) in the P region through
negative refraction. The response of a local probe (filled blue circle) in the
P region behaves as if the disturbance emanates not from the pump but instead
from the quantum mirage. (b) In a smooth P-N junction, the sharp quantum
mirage disappears, but its effect still exists. (c) Sketch of the electron
energy band in the N\ region and hole energy band in the P region. (d) Fermi
surface matching: the mirror reflection of the electron Fermi surface (red
curve) about the junction interface coincides with the hole Fermi surface
(blue curve).}%
\label{G_PNJ}%
\end{figure}

For an intuitive physical picture about the hidden quantum mirage, we start
from a sharp P-N junction as shown in Fig. \ref{G_PNJ}(a). Here the plane
waves excited by a local pump (filled red circle) in the N region is perfectly
focused by the junction into a sharp\textit{ }quantum mirage (red dashed
circle) in the P region \cite{CheianovScience2007,LeeNatPhys2015}.
Consequently, when a local probe (filled blue circle) is scanned in the P
region, the probe only \textquotedblleft sees\textquotedblright\ the mirage.
This motivates the natural expectation that the response of the probe would
behave as if the disturbance originated from this mirage instead of the pump.
When the interface becomes sufficiently smooth or disordered [see Fig.
\ref{G_PNJ}(b)], however, the quantum mirage disappears. Surprinsingly, the
pump-probe response is still independent of their distances when the pump and
probe undergo equal shifts in opposite directions perpendicular to the
interface, \textit{as if the quantum mirage were still there}.

To understand this \textit{robust}, \textit{hidden} quantum mirage effect,
from now on we specialize to P-N junctions (PNJs) of two-dimensional
semiconductors. The PNJ under consideration has translational invariance along
its interface [$y$ axis, see Fig. \ref{G_PNJ}(b)] and one electron (hole)
Fermi surface in the N (P) region [Figs. \ref{G_PNJ}(c) and (d)]. Apart from
this, the PNJ is arbitrary, e.g., the junction could be homo or hetero, sharp
or smooth, with arbitrary and even random potential profile. A right-going
plane wave $e^{i\mathbf{q}_{N}\cdot\mathbf{r}}$ with momentum $\mathbf{q}%
_{N}\equiv(q_{N,x},q_{y})$ ($q_{N,x}>0$) [red spot in Fig. \ref{G_PNJ}(c) and
(d)] on the electron Fermi surface goes across the junction and becomes a
right-going transmitted wave $e^{i\mathbf{q}_{P}\cdot\mathbf{r}}$ with
momentum $\mathbf{q}_{P}\equiv(q_{P,x},q_{y})$ ($q_{P,x}<0$) [blue spot in
Fig. \ref{G_PNJ}(c) and (d)] on the hole Fermi surface. The response amplitude
(as quantified by the retarded propagator on the Fermi surface) of the probe
located at $\mathbf{R}_{2}$ [filled blue circle in Fig. \ref{G_PNJ}(b)] in the
P region due to a local pump located at $\mathbf{R}_{1}$ [filled red circle in
Fig. \ref{G_PNJ}(b)] in the N region is dominated by these plane waves on the
Fermi surface (see supplementary online materials):%

\begin{equation}
{G}(\mathbf{R}_{2},\mathbf{R}_{1})=\int\frac{dq_{y}}{2\pi}w(q_{y}%
)e^{i(\mathbf{q}_{P}\cdot\mathbf{R}_{2}-\mathbf{q}_{N}\cdot\mathbf{R}_{1})},
\label{GF}%
\end{equation}
where $w(q_{y})$ is the transmission amplitude across the junction and
$e^{i(\mathbf{q}_{P}\cdot\mathbf{R}_{2}-\mathbf{q}_{N}\cdot\mathbf{R}_{1})}$
is the phase factor associated with the propagation outside the junction. Note
that the details of the junction (breadth, potential profile, etc) only
influences $w(q_{y})$ and has no effect on the propagation phase factor
outside the junction. For uniform systems, we have $\mathbf{q}_{N}%
=\mathbf{q}_{P}\equiv\mathbf{q}$, so the response only depends on the
displacement $\mathbf{R}_{2}-\mathbf{R}_{1}$ between the pump and the probe:
the rapid oscillation of the propagation phase factor $e^{i\mathbf{q}%
\cdot(\mathbf{R}_{2}-\mathbf{R}_{1})}$ at large distances leads to destructive
interference and hence $1/R^{1/2}$ decay of the response with the pump-probe
distance $R\equiv|\mathbf{R}_{2}-\mathbf{R}_{1}|$, which is just the $d=2$
case of the usual $1/R^{(d-1)/2}$ decay in a uniform $d$-dimensional
conducting system.

For PNJs, an interesting phenomenon appears when the mirror reflection of the
electron Fermi surface about the junction coincides with the hole Fermi
surface [Fig. \ref{G_PNJ}(d)], so that
\[
q_{N,x}=-q_{P,x}%
\]
for all $q_{y}$. For such Fermi-surface-matched PNJs, the propagation phase
factor becomes $e^{iq_{y}(Y_{2}-Y_{1})}e^{-iq_{N,x}(X_{2}+X_{1})}$, so the
response depends on the pump location $\mathbf{R}_{1}\equiv(X_{1}%
,Y_{1}\mathbf{)}$ and probe location $\mathbf{R}_{2}\equiv(X_{2}%
,Y_{2}\mathbf{)}$ through $Y_{2}-Y_{1}$ and $X_{2}+X_{1}$ only. In other
words, the response of the probe remains invariant not only upon identical
displacement of the pump and probe parallel to the junction (this invariance
trivially follows from the translational symmetry of the system along the
junction), but also upon opposite displacement of the pump and the probe
perpendicular to the junction (this invariance is absent from the Hamiltonian
and originates from Fermi surface matching), as if the disturbance emanated
\textit{not }from the pump but instead from its \textit{mirror point} about
the junction [dashed red circle in Fig. \ref{G_PNJ}(b)] although the
conventional quantum mirage
\cite{CheianovScience2007,ShalaevNatPhoton2007,JuzeliunasPRA2008,LederNatComm2014,LeeNatPhys2015}
already disappears.

The key to this hidden quantum mirage effect is that the momenta of the
propagating electrons (on the Fermi surface) in the N region and P region are
identical along the junction, but opposite perpendicular to the junction. As a
result, when the pump and probe are moved in opposite directions perpendicular
to the junction by an equal amount $\Delta X$, the extra propagation phase
factor $e^{iq_{P,x}\Delta X}$ in the P region is exactly cancelled by the
extra propagation phase factor $e^{iq_{N,x}\Delta X}$ in the N region. Even
when Fermi-surface matching is slightly broken, this physical picture could
still guarantee weak dependence of the nonlocal reponse on the pump-probe distance.

This hidden quantum mirage effect has two distinguishing features compared
with various conventional quantum mirages, such as electrostatic lens
\cite{SpectorAPL1990}, refocusing by an elliptical quantum corral
\cite{ManoharanNature2000}, and refocusing by negative refraction
\cite{CheianovScience2007,LederNatComm2014,LeeNatPhys2015}. First, it follows
entirely from symmetries (Fermi surface matching and 1D translational
invariance). Second, it cannot be directly observed as a mirage (i.e., local
enhancement of the probe response), but instead manifests itself as a
distance-independent response. This effect is applicable to either massless or
massive electrons in a wide range of materials. It also applies to other
matter waves and electromagnetic waves, thus the many experimental platforms
\cite{ParimiNature2003,FangScience2005,LiuScience2007,LezecScience2007,SoukoulisScience2007,XuNature2013,LeeNatPhys2015,LederNatComm2014}
that have been used for observing the refocusing by negative refraction for
various matter waves could be used to observe this more robust phenomenon.

Since the most popular 2D semiconductors have isotropic low-energy dispersion,
Fermi surface matching can be satisfied by appropriately tuning the Fermi
energy, so that the hidden quantum mirage can be detected by the
well-developed multiprobe scanning tunneling microscopy (STM), which has
already been used to characterize the non-local responses of many systems
\cite{LiAFM2013} such as two-dimensional thin films \cite{BannaniRSI2008} and
graphene \cite{SutterNatMater2008,JiNatMater2012} with nanoscale resolution.
We consider a two-dimensional PNJ connected to two STM tips, one located at
$\mathbf{R}_{1}$ in the N region and the other at $\mathbf{R}_{2}$ in the P
region. The zero-temperature conductance is given by the Landauer formula
\cite{DattaBook1995} as $G_{\mathrm{C}}=(2e^{2}/h)T,$ where
\[
T=|{G}(\mathbf{R}_{2},\mathbf{R}_{1})|^{2}\Gamma_{1}\Gamma_{2}%
\]
is the transmission coefficient and $\Gamma_{i}$ ($i=1,2$) is the coupling to
the $i$th probe. For uniform systems, the conductance decays as $1/R$ with
distance $R\equiv|\mathbf{R}_{2}-\mathbf{R}_{1}|$ due to the universal decay
${G}(\mathbf{R}_{2},\mathbf{R}_{1})\propto1/R^{1/2}$. In a Fermi-surface
matched PNJ, the conductance would be distance-independent when the two STM
tips are moved oppositely perpendicular to the junction. When the material has
significant spin splitting, e.g., due to intrinsic spin-orbit coupling or by
external exchange field from a ferromagnetic layer \cite{MoghaddamPRL2010},
the Fermi surface matching may occur for only one spin orientation, then the
hidden quantum mirage becomes spin-selective. This can be detected by
spin-polarized STM \cite{WiesendangerRMP2009,ZhouNatPhys2010}.

In addition to charge and spin transport, the hidden quantum mirage is also
applicable to carrier-mediated Rudermann-Kittel-Kasuya-Yosida (RKKY)
interaction between distant localized spins. In $d$-dimensional systems, the
RKKY interaction decays rapidly as $1/R^{d}$ with the inter-spin distance $R$
\cite{ZhuPRL2011}, limiting the spatial range of spin-spin correlation on 2D
systems that can be directly detected via spin-polarized scanning tunneling
microscopy to a few nanometers \cite{ZhouNatPhys2010,KhajetooriansNatPhys2012}%
. This universal decay of the RKKY interaction strength $J\sim G^{2}%
({\mathbf{R}_{2}},{\mathbf{R}_{1}})/R$ is a direct consequence of the
$1/R^{(d-1)/2}$ decay of the nonlocal response. Thus the hidden quantum mirage
could slow down the decay of the RKKY interaction from $1/R^{d}$ to $1/R$, and
hence drastically enhances its magnitude.

\begin{figure}[ptb]
\includegraphics[width=\columnwidth]{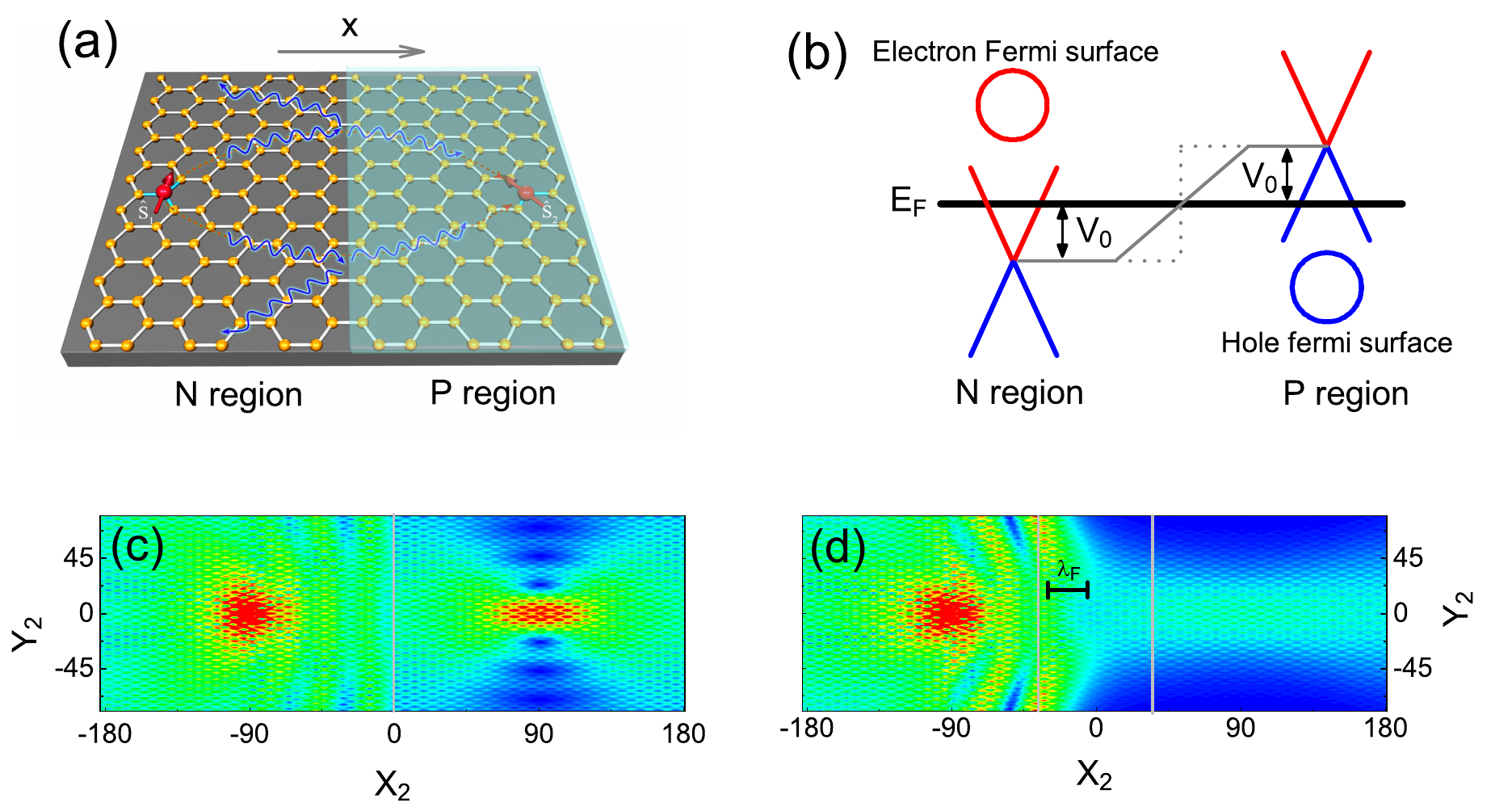}\caption{(color online).
(a) Graphene PNJ along the zigzag direction. (b) The gate voltages shift the
Dirac cone of the N (P) region by $-V_{0}$\ ($+V_{0}$). The N and P regions
are spatially separated by a sharp (green dashed line) or smooth linear (green
solid line) junction, whose interfaces are indicated by the vertical gray
lines. Fermi surface matching occurs when the Fermi level locates midway in
between the Dirac points, i.e., $E_{F}=0$. (c) shows the well-known quantum
mirage for a sharp junction: with a pump at $\mathbf{R}_{1}=(-91,0)$, the
response of a scanning probe $\mathbf{R}_{2}=(X_{2},Y_{2})$ shows a sharp
maximum at the mirror point $(+91,0)$. (d) shows that the quantum mirage
gradually disappears when the junction is wider than the Fermi wavelength
$\lambda_{F}$. In all the calculations $V_{0}=0.2$.}%
\label{G_GRAPHENEPNJ}%
\end{figure}

With the rapid progress in modern nanotechnology, high-quality PNJs have been
fabricated in graphene
\cite{YoungNatPhys2009,WilliamsNatNano2011,LeeNatPhys2015,RickhausNatComm2015}%
, transition-metal dichalcogenides
\cite{RossNatNano2014,BaugherNatNano2014,PospischilNatNano2014}, and are under
development for topological insulator surfaces
\cite{ZengAIPAdv2013,BathonArxiv2015}. We have performed extensive numerical
simulation that demonstrate the existence of hidden quantum mirage in these
systems (graphene, monolayer MoS$_{2}$, and topological insulator surfaces).
Here we present our results for graphene based on the tight-binding model. For
brevity, we use the lattice constant $a=1.42$ $%
%TCIMACRO{\unit{\r{A}}}%
%BeginExpansion
\operatorname{\r{A}}%
%EndExpansion
$ (nearest-neighbor hopping constant $t\approx3$ eV) of graphene as the unit
of length (energy).

\begin{figure}[ptb]
\includegraphics[width=\columnwidth]{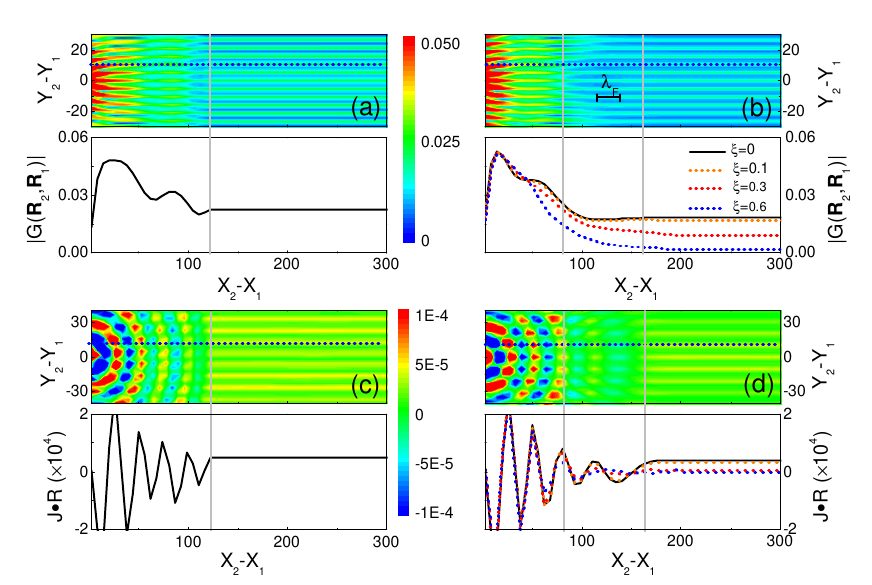}\caption{(color
online). (a) and (b): contour plot of pump-probe response amplitude
$|G(\mathbf{R}_{2},\mathbf{R}_{1})|$ vs. $\mathbf{R}_{2}-\mathbf{R}_{1}$ for
fixed $X_{2}+X_{1}=-120$ in (a)\ sharp or (b) smooth PNJ. A horizontal slice
(blue dashed line) of the contour plot is shown as the black solid lines in
the lower pannel. The dashed lines in the lower panel of (b) include on-site
disorder of different strengths $\xi$ in the smooth junction region along the
$x$ axis. (c) and (d): similar to (a) and (b), but for the scaled RKKY
interaction $JR$. In all the calculations $V_{0}=0.2$.}%
\label{G_SHORTCUT}%
\end{figure}

As illustrated in Fig. \ref{G_GRAPHENEPNJ}, voltages applied to the top and
back gates shift the electron Dirac cones down by $V_{0}$ on the left (N
region) and up by $V_{0}$ on the right (P region), so that Fermi surface
matching is achieved. As shown in Fig. \ref{G_SHORTCUT}(a) and (b), when we
move the pump and probe, initially at $\mathbf{R}_{1}=\mathbf{R}_{2}=(-60,0)$
in the N region, in opposite directions along the $x$ axis with $X_{2}+X_{1}$
fixed, the pump-probe response amplitude initially oscillates and decays, but
becomes invariant after the probe enters deep into the P region. Such
distance-independent response occurs not only in a sharp PNJ\ [Fig.
\ref{G_SHORTCUT}(a)], but also in a smooth PNJ with junction width $\gg$
carrier Fermi wavelength $\lambda_{F}$ [Fig. \ref{G_SHORTCUT}(b)], even when
on-site disorder of various strengths $\xi=0.1,$ $0.3,$ and $0.6$ along the
$x$ axis in the junction region has been introduced [dashed lines in the lower
pannel of Fig. \ref{G_SHORTCUT}(b)]. This demonstrates that except for the
translational symmetry and Fermi surface matching, the hidden quantum mirage
effect is robust against other details of the interface.

As a consequent of the distance-independent response, the carrier-mediated
RKKY\ interaction $J\hat{\mathbf{S}}_{1}\cdot\hat{\mathbf{S}}_{2}$ between two
magnetic moments $\hat{\mathbf{S}}_{1}$ (located at $\mathbf{R}_{1}$) and
$\hat{\mathbf{S}}_{2}$ (located at $\mathbf{R}_{2}$) is expected to decay with
inter-spin distance $R\equiv|\mathbf{R}_{2}-\mathbf{R}_{1}|$ as $J\propto1/R$.
Indeed, the scaled RKKY interaction strength $JR$ shows similar
distance-independent behaviors in both sharp [Fig. \ref{G_SHORTCUT}(c)] and
and smooth [Fig. \ref{G_SHORTCUT}(d)] PNJs, even in the presence of on-site
disorder [dashed lines in Fig. \ref{G_SHORTCUT}(d)]. This indicates\ the
robust $1/R$ decay of the RKKY interaction in graphene PNJs, as opposed to the
rapid $1/R^{2}$ decay in uniform graphene (or $1/R^{3}$ decay in uniform
undoped graphene). This $1/R$ scaling could dramatically amplify the
RKKY\ interaction in graphene PNJs compared with that in uniform graphene.

\begin{figure}[ptb]
\includegraphics[width=0.8\columnwidth]{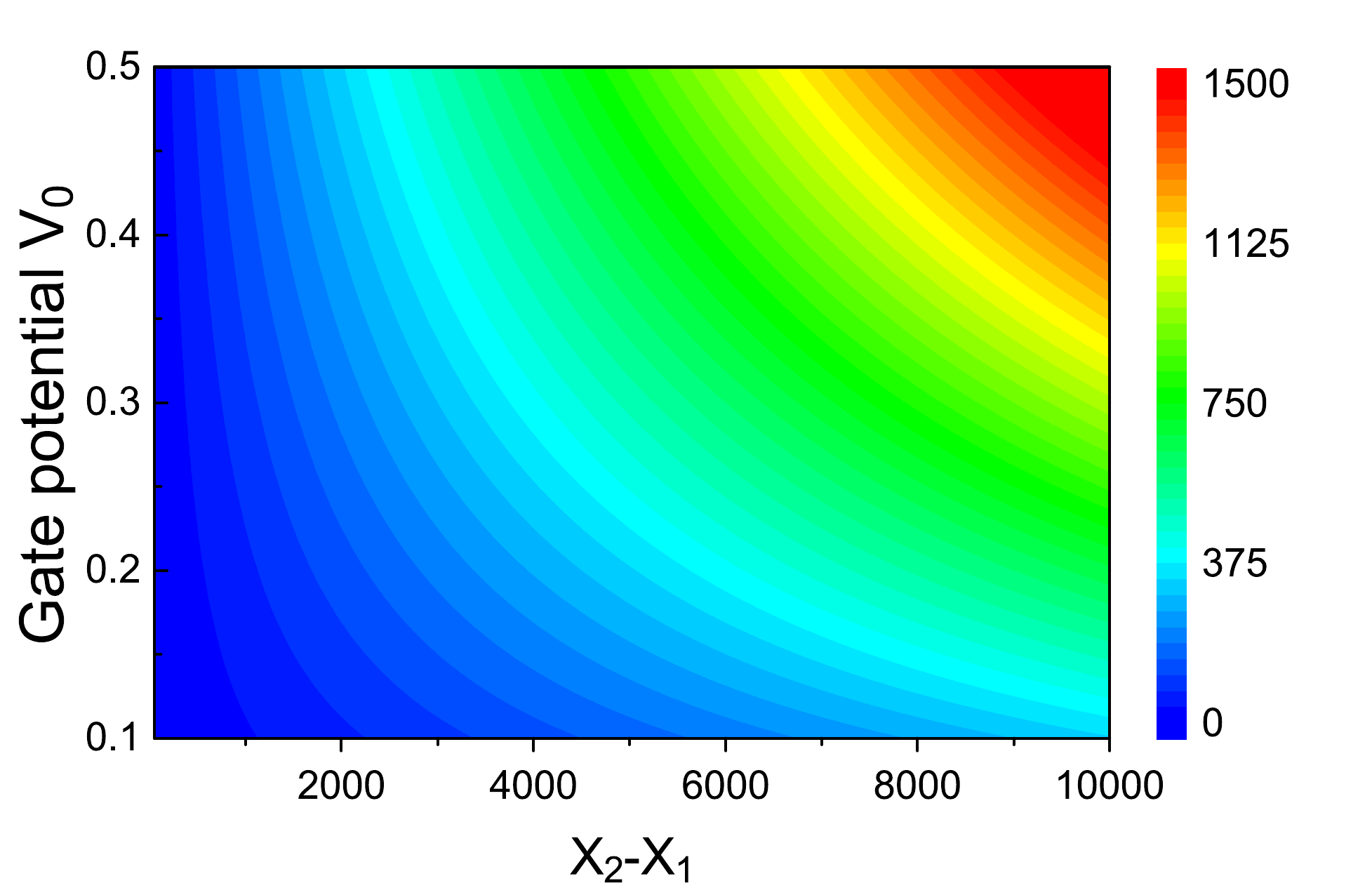}\caption{(color
online). Amplification factor of RKKY interaction vs. gate voltage and
spin-spin distance.}%
\label{G_AMPLIFY}%
\end{figure}

Maximal amplification is achieved when the hidden quantum mirage becomes
visible in a sharp PNJ satisfying Fermi surface matching. The RKKY interaction
attains its maximum value $J_{\max}$ when the second magnetic moment locates
at the quantum mirage of the first magnetic moment. Extensive numerical
simulation for different $\mathbf{R}_{1},\mathbf{R}_{2}$ and junction
potential $V_{0}$ shows that in the linear dispersion regime ($V_{0}\leq1$)
and for long inter-spin distances $R\gg\lambda_{F}$, the maximal RKKY
interaction $J_{\max}\approx CV_{0}^{2}/R$, where $C$ is a constant. For
comparison, in uniform graphene with the same carrier concentration (i.e.,
Fermi energy $|E_{F}|=V_{0}$), the magnitude of the RKKY interaction
$J_{0}\approx3CV_{0}/R$. Therefore, the PNJ amplifies the RKKY interaction by
a factor
\[
\eta=\frac{J_{\max}}{J_{0}}\approx\frac{V_{0}R}{3}%
\]
that could reach three orders of magnitudes (Fig. \ref{G_AMPLIFY}). For a
typical junction potential $V_{0}=1$ eV \cite{WilliamsNatNano2011} and two
substitutional manganese spins separated by 16 nm, their RKKY $J\sim16$
$\mathrm{\mu eV}$ (amplified by a factor $\eta\approx13$) is measurable by
spin-polarized scanning tunneling spectroscopy \cite{ZhouNatPhys2010}. When
the distance increases up to the ballistic length $\sim1$ $\mathrm{\mu m}$
\cite{RickhausNatComm2015}, the RKKY$\ $interaction $J\sim0.3$ $\mathrm{\mu
eV}$ (amplified by a factor $\eta\approx780$) may be detected by an
ultrasensitive magnetic sensor based on nitrogen-vacancy center in
nanodiamonds \cite{MazeNature2008,TaylorNatPhys2008}, which has demonstrated
nanoscale spatial resolution and the capability to determine weak magnetic
dipolar interaction $\sim10^{-5}$ $\mathrm{\mu eV}$ between two nuclear spins
\cite{ZhaoNatNano2011,ShiNatPhys2014}.

In summary, we have proposed a robust, hidden quantum mirage that could
dramatically enhance the non-local responses of electrons as well as the
carrier-mediated interaction. This effect also applies to electromagnetic
waves and other matter waves such as cold atoms. It may be useful for
engineering energy, charge, and spin transport as well as carrier-mediated
long-range correlation.

This work was supported by the MOST (Grant No. 2015CB921503, and No.
2014CB848700) and NSFC (Grant No. 11434010, No. 11274036, No. 11322542, and
No. 11404043). J. J. Z. thanks the new research direction support program of
CQUPT. We acknowledge the computational support from the Beijing Computational
Science Research Center (CSRC).
%\bibliographystyle{apsrev4-1}
%\bibliography{e://dropbox/MyLiterature}

\begin{thebibliography}{50}%
\makeatletter
\providecommand \@ifxundefined [1]{%
 \@ifx{#1\undefined}
}%
\providecommand \@ifnum [1]{%
 \ifnum #1\expandafter \@firstoftwo
 \else \expandafter \@secondoftwo
 \fi
}%
\providecommand \@ifx [1]{%
 \ifx #1\expandafter \@firstoftwo
 \else \expandafter \@secondoftwo
 \fi
}%
\providecommand \natexlab [1]{#1}%
\providecommand \enquote  [1]{``#1''}%
\providecommand \bibnamefont  [1]{#1}%
\providecommand \bibfnamefont [1]{#1}%
\providecommand \citenamefont [1]{#1}%
\providecommand \href@noop [0]{\@secondoftwo}%
\providecommand \href [0]{\begingroup \@sanitize@url \@href}%
\providecommand \@href[1]{\@@startlink{#1}\@@href}%
\providecommand \@@href[1]{\endgroup#1\@@endlink}%
\providecommand \@sanitize@url [0]{\catcode `\\12\catcode `\$12\catcode
  `\&12\catcode `\#12\catcode `\^12\catcode `\_12\catcode `\%12\relax}%
\providecommand \@@startlink[1]{}%
\providecommand \@@endlink[0]{}%
\providecommand \url  [0]{\begingroup\@sanitize@url \@url }%
\providecommand \@url [1]{\endgroup\@href {#1}{\urlprefix }}%
\providecommand \urlprefix  [0]{URL }%
\providecommand \Eprint [0]{\href }%
\providecommand \doibase [0]{http://dx.doi.org/}%
\providecommand \selectlanguage [0]{\@gobble}%
\providecommand \bibinfo  [0]{\@secondoftwo}%
\providecommand \bibfield  [0]{\@secondoftwo}%
\providecommand \translation [1]{[#1]}%
\providecommand \BibitemOpen [0]{}%
\providecommand \bibitemStop [0]{}%
\providecommand \bibitemNoStop [0]{.\EOS\space}%
\providecommand \EOS [0]{\spacefactor3000\relax}%
\providecommand \BibitemShut  [1]{\csname bibitem#1\endcsname}%
\let\auto@bib@innerbib\@empty
%</preamble>
\bibitem [{\citenamefont {Veselago}(1968)}]{VeselagoSPU1968}%
  \BibitemOpen
  \bibfield  {author} {\bibinfo {author} {\bibfnamefont {V.~G.}\ \bibnamefont
  {Veselago}},\ }\href@noop {} {\bibfield  {journal} {\bibinfo  {journal}
  {Soviet Physics Uspekhi}\ }\textbf {\bibinfo {volume} {10}},\ \bibinfo
  {pages} {509} (\bibinfo {year} {1968})}\BibitemShut {NoStop}%
\bibitem [{\citenamefont {Pendry}(2000)}]{PendryPRL2000}%
  \BibitemOpen
  \bibfield  {author} {\bibinfo {author} {\bibfnamefont {J.~B.}\ \bibnamefont
  {Pendry}},\ }\href@noop {} {\bibfield  {journal} {\bibinfo  {journal} {Phys.
  Rev. Lett.}\ }\textbf {\bibinfo {volume} {85}},\ \bibinfo {pages} {3966}
  (\bibinfo {year} {2000})}\BibitemShut {NoStop}%
\bibitem [{\citenamefont {Zhang}\ and\ \citenamefont
  {Liu}(2008)}]{ZhangNatMater2008}%
  \BibitemOpen
  \bibfield  {author} {\bibinfo {author} {\bibfnamefont {X.}~\bibnamefont
  {Zhang}}\ and\ \bibinfo {author} {\bibfnamefont {Z.}~\bibnamefont {Liu}},\
  }\href@noop {} {\bibfield  {journal} {\bibinfo  {journal} {Nat Mater}\
  }\textbf {\bibinfo {volume} {7}},\ \bibinfo {pages} {435} (\bibinfo {year}
  {2008})}\BibitemShut {NoStop}%
\bibitem [{\citenamefont {Pendry}\ \emph {et~al.}(2012)\citenamefont {Pendry},
  \citenamefont {Aubry}, \citenamefont {Smith},\ and\ \citenamefont
  {Maier}}]{PendryScience2012}%
  \BibitemOpen
  \bibfield  {author} {\bibinfo {author} {\bibfnamefont {J.~B.}\ \bibnamefont
  {Pendry}}, \bibinfo {author} {\bibfnamefont {A.}~\bibnamefont {Aubry}},
  \bibinfo {author} {\bibfnamefont {D.~R.}\ \bibnamefont {Smith}}, \ and\
  \bibinfo {author} {\bibfnamefont {S.~A.}\ \bibnamefont {Maier}},\ }\href@noop
  {} {\bibfield  {journal} {\bibinfo  {journal} {Science}\ }\textbf {\bibinfo
  {volume} {337}},\ \bibinfo {pages} {549} (\bibinfo {year}
  {2012})}\BibitemShut {NoStop}%
\bibitem [{\citenamefont {Manoharan}\ \emph {et~al.}(2000)\citenamefont
  {Manoharan}, \citenamefont {Lutz},\ and\ \citenamefont
  {Eigler}}]{ManoharanNature2000}%
  \BibitemOpen
  \bibfield  {author} {\bibinfo {author} {\bibfnamefont {H.~C.}\ \bibnamefont
  {Manoharan}}, \bibinfo {author} {\bibfnamefont {C.~P.}\ \bibnamefont {Lutz}},
  \ and\ \bibinfo {author} {\bibfnamefont {D.~M.}\ \bibnamefont {Eigler}},\
  }\href@noop {} {\bibfield  {journal} {\bibinfo  {journal} {Nature}\ }\textbf
  {\bibinfo {volume} {403}},\ \bibinfo {pages} {512} (\bibinfo {year}
  {2000})}\BibitemShut {NoStop}%
\bibitem [{\citenamefont {Shalaev}(2007)}]{ShalaevNatPhoton2007}%
  \BibitemOpen
  \bibfield  {author} {\bibinfo {author} {\bibfnamefont {V.~M.}\ \bibnamefont
  {Shalaev}},\ }\href@noop {} {\bibfield  {journal} {\bibinfo  {journal} {Nat
  Photon}\ }\textbf {\bibinfo {volume} {1}},\ \bibinfo {pages} {41} (\bibinfo
  {year} {2007})}\BibitemShut {NoStop}%
\bibitem [{\citenamefont {Juzeliunas}\ \emph {et~al.}(2008)\citenamefont
  {Juzeliunas}, \citenamefont {Ruseckas}, \citenamefont {Lindberg},
  \citenamefont {Santos},\ and\ \citenamefont {\"Ohberg}}]{JuzeliunasPRA2008}%
  \BibitemOpen
  \bibfield  {author} {\bibinfo {author} {\bibfnamefont {G.}~\bibnamefont
  {Juzeliunas}}, \bibinfo {author} {\bibfnamefont {J.}~\bibnamefont
  {Ruseckas}}, \bibinfo {author} {\bibfnamefont {M.}~\bibnamefont {Lindberg}},
  \bibinfo {author} {\bibfnamefont {L.}~\bibnamefont {Santos}}, \ and\ \bibinfo
  {author} {\bibfnamefont {P.}~\bibnamefont {\"Ohberg}},\ }\href@noop {}
  {\bibfield  {journal} {\bibinfo  {journal} {Phys. Rev. A}\ }\textbf {\bibinfo
  {volume} {77}},\ \bibinfo {pages} {011802} (\bibinfo {year}
  {2008})}\BibitemShut {NoStop}%
\bibitem [{\citenamefont {Leder}\ \emph {et~al.}(2014)\citenamefont {Leder},
  \citenamefont {Grossert},\ and\ \citenamefont {Weitz}}]{LederNatComm2014}%
  \BibitemOpen
  \bibfield  {author} {\bibinfo {author} {\bibfnamefont {M.}~\bibnamefont
  {Leder}}, \bibinfo {author} {\bibfnamefont {C.}~\bibnamefont {Grossert}}, \
  and\ \bibinfo {author} {\bibfnamefont {M.}~\bibnamefont {Weitz}},\
  }\href@noop {} {\bibfield  {journal} {\bibinfo  {journal} {Nat Commun}\
  }\textbf {\bibinfo {volume} {5}},\ \bibinfo {pages} {3327} (\bibinfo {year}
  {2014})}\BibitemShut {NoStop}%
\bibitem [{\citenamefont {Cheianov}\ \emph {et~al.}(2007)\citenamefont
  {Cheianov}, \citenamefont {Fal'ko},\ and\ \citenamefont
  {Altshuler}}]{CheianovScience2007}%
  \BibitemOpen
  \bibfield  {author} {\bibinfo {author} {\bibfnamefont {V.~V.}\ \bibnamefont
  {Cheianov}}, \bibinfo {author} {\bibfnamefont {V.}~\bibnamefont {Fal'ko}}, \
  and\ \bibinfo {author} {\bibfnamefont {B.~L.}\ \bibnamefont {Altshuler}},\
  }\href@noop {} {\bibfield  {journal} {\bibinfo  {journal} {Science}\ }\textbf
  {\bibinfo {volume} {315}},\ \bibinfo {pages} {1252} (\bibinfo {year}
  {2007})}\BibitemShut {NoStop}%
\bibitem [{\citenamefont {Park}\ \emph {et~al.}(2008)\citenamefont {Park},
  \citenamefont {Son}, \citenamefont {Yang}, \citenamefont {Cohen},\ and\
  \citenamefont {Louie}}]{ParkNanoLett2008}%
  \BibitemOpen
  \bibfield  {author} {\bibinfo {author} {\bibfnamefont {C.-H.}\ \bibnamefont
  {Park}}, \bibinfo {author} {\bibfnamefont {Y.-W.}\ \bibnamefont {Son}},
  \bibinfo {author} {\bibfnamefont {L.}~\bibnamefont {Yang}}, \bibinfo {author}
  {\bibfnamefont {M.~L.}\ \bibnamefont {Cohen}}, \ and\ \bibinfo {author}
  {\bibfnamefont {S.~G.}\ \bibnamefont {Louie}},\ }\href@noop {} {\bibfield
  {journal} {\bibinfo  {journal} {Nano Lett.}\ }\textbf {\bibinfo {volume}
  {8}},\ \bibinfo {pages} {2920} (\bibinfo {year} {2008})}\BibitemShut
  {NoStop}%
\bibitem [{\citenamefont {Moghaddam}\ and\ \citenamefont
  {Zareyan}(2010)}]{MoghaddamPRL2010}%
  \BibitemOpen
  \bibfield  {author} {\bibinfo {author} {\bibfnamefont {A.~G.}\ \bibnamefont
  {Moghaddam}}\ and\ \bibinfo {author} {\bibfnamefont {M.}~\bibnamefont
  {Zareyan}},\ }\href@noop {} {\bibfield  {journal} {\bibinfo  {journal} {Phys.
  Rev. Lett.}\ }\textbf {\bibinfo {volume} {105}},\ \bibinfo {pages} {146803}
  (\bibinfo {year} {2010})}\BibitemShut {NoStop}%
\bibitem [{\citenamefont {Zhao}\ \emph {et~al.}(2013)\citenamefont {Zhao},
  \citenamefont {Tang}, \citenamefont {Gu},\ and\ \citenamefont
  {Duan}}]{ZhaoPRL2013}%
  \BibitemOpen
  \bibfield  {author} {\bibinfo {author} {\bibfnamefont {L.}~\bibnamefont
  {Zhao}}, \bibinfo {author} {\bibfnamefont {P.}~\bibnamefont {Tang}}, \bibinfo
  {author} {\bibfnamefont {B.-L.}\ \bibnamefont {Gu}}, \ and\ \bibinfo {author}
  {\bibfnamefont {W.}~\bibnamefont {Duan}},\ }\href@noop {} {\bibfield
  {journal} {\bibinfo  {journal} {Phys. Rev. Lett.}\ }\textbf {\bibinfo
  {volume} {111}},\ \bibinfo {pages} {116601} (\bibinfo {year}
  {2013})}\BibitemShut {NoStop}%
\bibitem [{\citenamefont {Lee}\ \emph {et~al.}(2015)\citenamefont {Lee},
  \citenamefont {Park},\ and\ \citenamefont {Lee}}]{LeeNatPhys2015}%
  \BibitemOpen
  \bibfield  {author} {\bibinfo {author} {\bibfnamefont {G.-H.}\ \bibnamefont
  {Lee}}, \bibinfo {author} {\bibfnamefont {G.-H.}\ \bibnamefont {Park}}, \
  and\ \bibinfo {author} {\bibfnamefont {H.-J.}\ \bibnamefont {Lee}},\
  }\href@noop {} {\bibfield  {journal} {\bibinfo  {journal} {Nat. Phys.}\
  }\textbf {\bibinfo {volume} {11}},\ \bibinfo {pages} {925} (\bibinfo {year}
  {2015})}\BibitemShut {NoStop}%
\bibitem [{\citenamefont {Igor}\ \emph {et~al.}(2004)\citenamefont {Igor},
  \citenamefont {Fabian},\ and\ \citenamefont {Das~Sarma}}]{IgorRMP2004}%
  \BibitemOpen
  \bibfield  {author} {\bibinfo {author} {\bibfnamefont {Z.}~\bibnamefont
  {Igor}}, \bibinfo {author} {\bibfnamefont {J.}~\bibnamefont {Fabian}}, \ and\
  \bibinfo {author} {\bibfnamefont {S.}~\bibnamefont {Das~Sarma}},\ }\href@noop
  {} {\bibfield  {journal} {\bibinfo  {journal} {Rev. Mod. Phys.}\ }\textbf
  {\bibinfo {volume} {76}},\ \bibinfo {pages} {323} (\bibinfo {year}
  {2004})}\BibitemShut {NoStop}%
\bibitem [{\citenamefont {Cirac¨¬}\ \emph {et~al.}(2012)\citenamefont
  {Cirac¨¬}, \citenamefont {Hill}, \citenamefont {Mock}, \citenamefont
  {Urzhumov}, \citenamefont {Fern¨¢ndez-Dom¨ªnguez}, \citenamefont {Maier},
  \citenamefont {Pendry}, \citenamefont {Chilkoti},\ and\ \citenamefont
  {Smith}}]{CiraciScience2012}%
  \BibitemOpen
  \bibfield  {author} {\bibinfo {author} {\bibfnamefont {C.}~\bibnamefont
  {Cirac¨¬}}, \bibinfo {author} {\bibfnamefont {R.~T.}\ \bibnamefont {Hill}},
  \bibinfo {author} {\bibfnamefont {J.~J.}\ \bibnamefont {Mock}}, \bibinfo
  {author} {\bibfnamefont {Y.}~\bibnamefont {Urzhumov}}, \bibinfo {author}
  {\bibfnamefont {A.~I.}\ \bibnamefont {Fern¨¢ndez-Dom¨ªnguez}}, \bibinfo
  {author} {\bibfnamefont {S.~A.}\ \bibnamefont {Maier}}, \bibinfo {author}
  {\bibfnamefont {J.~B.}\ \bibnamefont {Pendry}}, \bibinfo {author}
  {\bibfnamefont {A.}~\bibnamefont {Chilkoti}}, \ and\ \bibinfo {author}
  {\bibfnamefont {D.~R.}\ \bibnamefont {Smith}},\ }\href@noop {} {\bibfield
  {journal} {\bibinfo  {journal} {Science}\ }\textbf {\bibinfo {volume}
  {337}},\ \bibinfo {pages} {1072} (\bibinfo {year} {2012})}\BibitemShut
  {NoStop}%
\bibitem [{\citenamefont {Cheianov}\ and\ \citenamefont
  {Fal'ko}(2006)}]{CheianovPRL2006}%
  \BibitemOpen
  \bibfield  {author} {\bibinfo {author} {\bibfnamefont {V.~V.}\ \bibnamefont
  {Cheianov}}\ and\ \bibinfo {author} {\bibfnamefont {V.~I.}\ \bibnamefont
  {Fal'ko}},\ }\href@noop {} {\bibfield  {journal} {\bibinfo  {journal} {Phys.
  Rev. Lett.}\ }\textbf {\bibinfo {volume} {97}},\ \bibinfo {pages} {226801}
  (\bibinfo {year} {2006})}\BibitemShut {NoStop}%
\bibitem [{\citenamefont {Hwang}\ and\ \citenamefont
  {Das~Sarma}(2008)}]{HwangPRL2008}%
  \BibitemOpen
  \bibfield  {author} {\bibinfo {author} {\bibfnamefont {E.~H.}\ \bibnamefont
  {Hwang}}\ and\ \bibinfo {author} {\bibfnamefont {S.}~\bibnamefont
  {Das~Sarma}},\ }\href@noop {} {\bibfield  {journal} {\bibinfo  {journal}
  {Phys. Rev. Lett.}\ }\textbf {\bibinfo {volume} {101}},\ \bibinfo {pages}
  {156802} (\bibinfo {year} {2008})}\BibitemShut {NoStop}%
\bibitem [{\citenamefont {Bena}(2008)}]{BenaPRL2008}%
  \BibitemOpen
  \bibfield  {author} {\bibinfo {author} {\bibfnamefont {C.}~\bibnamefont
  {Bena}},\ }\href@noop {} {\bibfield  {journal} {\bibinfo  {journal} {Phys.
  Rev. Lett.}\ }\textbf {\bibinfo {volume} {100}},\ \bibinfo {pages} {076601}
  (\bibinfo {year} {2008})}\BibitemShut {NoStop}%
\bibitem [{\citenamefont {Ruderman}\ and\ \citenamefont
  {Kittel}(1954)}]{RudermanPR1954}%
  \BibitemOpen
  \bibfield  {author} {\bibinfo {author} {\bibfnamefont {M.~A.}\ \bibnamefont
  {Ruderman}}\ and\ \bibinfo {author} {\bibfnamefont {C.}~\bibnamefont
  {Kittel}},\ }\href@noop {} {\bibfield  {journal} {\bibinfo  {journal} {Phys.
  Rev.}\ }\textbf {\bibinfo {volume} {96}},\ \bibinfo {pages} {99} (\bibinfo
  {year} {1954})}\BibitemShut {NoStop}%
\bibitem [{\citenamefont {Kasuya}(1956)}]{KasuyaPTP1956}%
  \BibitemOpen
  \bibfield  {author} {\bibinfo {author} {\bibfnamefont {T.}~\bibnamefont
  {Kasuya}},\ }\href@noop {} {\bibfield  {journal} {\bibinfo  {journal} {Prog.
  Theor. Phys.}\ }\textbf {\bibinfo {volume} {16}},\ \bibinfo {pages} {45}
  (\bibinfo {year} {1956})}\BibitemShut {NoStop}%
\bibitem [{\citenamefont {Yosida}(1957)}]{YosidaPR1957}%
  \BibitemOpen
  \bibfield  {author} {\bibinfo {author} {\bibfnamefont {K.}~\bibnamefont
  {Yosida}},\ }\href@noop {} {\bibfield  {journal} {\bibinfo  {journal} {Phys.
  Rev.}\ }\textbf {\bibinfo {volume} {106}},\ \bibinfo {pages} {893} (\bibinfo
  {year} {1957})}\BibitemShut {NoStop}%
\bibitem [{\citenamefont {Dietl}\ and\ \citenamefont
  {Ohno}(2014)}]{DietlRMP2014}%
  \BibitemOpen
  \bibfield  {author} {\bibinfo {author} {\bibfnamefont {T.}~\bibnamefont
  {Dietl}}\ and\ \bibinfo {author} {\bibfnamefont {H.}~\bibnamefont {Ohno}},\
  }\href@noop {} {\bibfield  {journal} {\bibinfo  {journal} {Rev. Mod. Phys.}\
  }\textbf {\bibinfo {volume} {86}},\ \bibinfo {pages} {187} (\bibinfo {year}
  {2014})}\BibitemShut {NoStop}%
\bibitem [{\citenamefont {Spector}\ \emph {et~al.}(1990)\citenamefont
  {Spector}, \citenamefont {Stormer}, \citenamefont {Baldwin}, \citenamefont
  {Pfeiffer},\ and\ \citenamefont {West}}]{SpectorAPL1990}%
  \BibitemOpen
  \bibfield  {author} {\bibinfo {author} {\bibfnamefont {J.}~\bibnamefont
  {Spector}}, \bibinfo {author} {\bibfnamefont {H.~L.}\ \bibnamefont
  {Stormer}}, \bibinfo {author} {\bibfnamefont {K.~W.}\ \bibnamefont
  {Baldwin}}, \bibinfo {author} {\bibfnamefont {L.~N.}\ \bibnamefont
  {Pfeiffer}}, \ and\ \bibinfo {author} {\bibfnamefont {K.~W.}\ \bibnamefont
  {West}},\ }\href@noop {} {\bibfield  {journal} {\bibinfo  {journal} {Appl.
  Phys. Lett.}\ }\textbf {\bibinfo {volume} {56}},\ \bibinfo {pages} {1290}
  (\bibinfo {year} {1990})}\BibitemShut {NoStop}%
\bibitem [{\citenamefont {Parimi}\ \emph {et~al.}(2003)\citenamefont {Parimi},
  \citenamefont {Lu}, \citenamefont {Vodo},\ and\ \citenamefont
  {Sridhar}}]{ParimiNature2003}%
  \BibitemOpen
  \bibfield  {author} {\bibinfo {author} {\bibfnamefont {P.~V.}\ \bibnamefont
  {Parimi}}, \bibinfo {author} {\bibfnamefont {W.~T.}\ \bibnamefont {Lu}},
  \bibinfo {author} {\bibfnamefont {P.}~\bibnamefont {Vodo}}, \ and\ \bibinfo
  {author} {\bibfnamefont {S.}~\bibnamefont {Sridhar}},\ }\href@noop {}
  {\bibfield  {journal} {\bibinfo  {journal} {Nature}\ }\textbf {\bibinfo
  {volume} {426}},\ \bibinfo {pages} {404} (\bibinfo {year}
  {2003})}\BibitemShut {NoStop}%
\bibitem [{\citenamefont {Fang}\ \emph {et~al.}(2005)\citenamefont {Fang},
  \citenamefont {Lee}, \citenamefont {Sun},\ and\ \citenamefont
  {Zhang}}]{FangScience2005}%
  \BibitemOpen
  \bibfield  {author} {\bibinfo {author} {\bibfnamefont {N.}~\bibnamefont
  {Fang}}, \bibinfo {author} {\bibfnamefont {H.}~\bibnamefont {Lee}}, \bibinfo
  {author} {\bibfnamefont {C.}~\bibnamefont {Sun}}, \ and\ \bibinfo {author}
  {\bibfnamefont {X.}~\bibnamefont {Zhang}},\ }\href@noop {} {\bibfield
  {journal} {\bibinfo  {journal} {Science}\ }\textbf {\bibinfo {volume}
  {308}},\ \bibinfo {pages} {534} (\bibinfo {year} {2005})}\BibitemShut
  {NoStop}%
\bibitem [{\citenamefont {Liu}\ \emph {et~al.}(2007)\citenamefont {Liu},
  \citenamefont {Lee}, \citenamefont {Xiong}, \citenamefont {Sun},\ and\
  \citenamefont {Zhang}}]{LiuScience2007}%
  \BibitemOpen
  \bibfield  {author} {\bibinfo {author} {\bibfnamefont {Z.}~\bibnamefont
  {Liu}}, \bibinfo {author} {\bibfnamefont {H.}~\bibnamefont {Lee}}, \bibinfo
  {author} {\bibfnamefont {Y.}~\bibnamefont {Xiong}}, \bibinfo {author}
  {\bibfnamefont {C.}~\bibnamefont {Sun}}, \ and\ \bibinfo {author}
  {\bibfnamefont {X.}~\bibnamefont {Zhang}},\ }\href@noop {} {\bibfield
  {journal} {\bibinfo  {journal} {Science}\ }\textbf {\bibinfo {volume}
  {315}},\ \bibinfo {pages} {1686} (\bibinfo {year} {2007})}\BibitemShut
  {NoStop}%
\bibitem [{\citenamefont {Lezec}\ \emph {et~al.}(2007)\citenamefont {Lezec},
  \citenamefont {Dionne},\ and\ \citenamefont {Atwater}}]{LezecScience2007}%
  \BibitemOpen
  \bibfield  {author} {\bibinfo {author} {\bibfnamefont {H.~J.}\ \bibnamefont
  {Lezec}}, \bibinfo {author} {\bibfnamefont {J.~A.}\ \bibnamefont {Dionne}}, \
  and\ \bibinfo {author} {\bibfnamefont {H.~A.}\ \bibnamefont {Atwater}},\
  }\href@noop {} {\bibfield  {journal} {\bibinfo  {journal} {Science}\ }\textbf
  {\bibinfo {volume} {316}},\ \bibinfo {pages} {430} (\bibinfo {year}
  {2007})}\BibitemShut {NoStop}%
\bibitem [{\citenamefont {Soukoulis}\ \emph {et~al.}(2007)\citenamefont
  {Soukoulis}, \citenamefont {Linden},\ and\ \citenamefont
  {Wegener}}]{SoukoulisScience2007}%
  \BibitemOpen
  \bibfield  {author} {\bibinfo {author} {\bibfnamefont {C.~M.}\ \bibnamefont
  {Soukoulis}}, \bibinfo {author} {\bibfnamefont {S.}~\bibnamefont {Linden}}, \
  and\ \bibinfo {author} {\bibfnamefont {M.}~\bibnamefont {Wegener}},\
  }\href@noop {} {\bibfield  {journal} {\bibinfo  {journal} {Science}\ }\textbf
  {\bibinfo {volume} {315}},\ \bibinfo {pages} {47} (\bibinfo {year}
  {2007})}\BibitemShut {NoStop}%
\bibitem [{\citenamefont {Xu}\ \emph {et~al.}(2013)\citenamefont {Xu},
  \citenamefont {Agrawal}, \citenamefont {Abashin}, \citenamefont {Chau},\ and\
  \citenamefont {Lezec}}]{XuNature2013}%
  \BibitemOpen
  \bibfield  {author} {\bibinfo {author} {\bibfnamefont {T.}~\bibnamefont
  {Xu}}, \bibinfo {author} {\bibfnamefont {A.}~\bibnamefont {Agrawal}},
  \bibinfo {author} {\bibfnamefont {M.}~\bibnamefont {Abashin}}, \bibinfo
  {author} {\bibfnamefont {K.~J.}\ \bibnamefont {Chau}}, \ and\ \bibinfo
  {author} {\bibfnamefont {H.~J.}\ \bibnamefont {Lezec}},\ }\href@noop {}
  {\bibfield  {journal} {\bibinfo  {journal} {Nature}\ }\textbf {\bibinfo
  {volume} {497}},\ \bibinfo {pages} {470} (\bibinfo {year}
  {2013})}\BibitemShut {NoStop}%
\bibitem [{\citenamefont {Li}\ \emph {et~al.}(2013)\citenamefont {Li},
  \citenamefont {Clark}, \citenamefont {Zhang},\ and\ \citenamefont
  {Baddorf}}]{LiAFM2013}%
  \BibitemOpen
  \bibfield  {author} {\bibinfo {author} {\bibfnamefont {A.-P.}\ \bibnamefont
  {Li}}, \bibinfo {author} {\bibfnamefont {K.~W.}\ \bibnamefont {Clark}},
  \bibinfo {author} {\bibfnamefont {X.-G.}\ \bibnamefont {Zhang}}, \ and\
  \bibinfo {author} {\bibfnamefont {A.~P.}\ \bibnamefont {Baddorf}},\
  }\href@noop {} {\bibfield  {journal} {\bibinfo  {journal} {Advanced
  Functional Materials}\ }\textbf {\bibinfo {volume} {23}},\ \bibinfo {pages}
  {2509} (\bibinfo {year} {2013})}\BibitemShut {NoStop}%
\bibitem [{\citenamefont {Bannani}\ \emph {et~al.}(2008)\citenamefont
  {Bannani}, \citenamefont {Bobisch},\ and\ \citenamefont
  {Möller}}]{BannaniRSI2008}%
  \BibitemOpen
  \bibfield  {author} {\bibinfo {author} {\bibfnamefont {A.}~\bibnamefont
  {Bannani}}, \bibinfo {author} {\bibfnamefont {C.~A.}\ \bibnamefont
  {Bobisch}}, \ and\ \bibinfo {author} {\bibfnamefont {R.}~\bibnamefont
  {Möller}},\ }\href@noop {} {\bibfield  {journal} {\bibinfo  {journal}
  {Review of Scientific Instruments}\ }\textbf {\bibinfo {volume} {79}},\
  \bibinfo {eid} {083704} (\bibinfo {year} {2008})}\BibitemShut {NoStop}%
\bibitem [{\citenamefont {Sutter}\ \emph {et~al.}(2008)\citenamefont {Sutter},
  \citenamefont {Flege},\ and\ \citenamefont {Sutter}}]{SutterNatMater2008}%
  \BibitemOpen
  \bibfield  {author} {\bibinfo {author} {\bibfnamefont {P.~W.}\ \bibnamefont
  {Sutter}}, \bibinfo {author} {\bibfnamefont {J.-I.}\ \bibnamefont {Flege}}, \
  and\ \bibinfo {author} {\bibfnamefont {E.~A.}\ \bibnamefont {Sutter}},\
  }\href@noop {} {\bibfield  {journal} {\bibinfo  {journal} {Nat Mater}\
  }\textbf {\bibinfo {volume} {7}},\ \bibinfo {pages} {406} (\bibinfo {year}
  {2008})}\BibitemShut {NoStop}%
\bibitem [{\citenamefont {Ji}\ \emph {et~al.}(2012)\citenamefont {Ji},
  \citenamefont {Hannon}, \citenamefont {Tromp}, \citenamefont {Perebeinos},
  \citenamefont {Tersoff},\ and\ \citenamefont {Ross}}]{JiNatMater2012}%
  \BibitemOpen
  \bibfield  {author} {\bibinfo {author} {\bibfnamefont {S.-H.}\ \bibnamefont
  {Ji}}, \bibinfo {author} {\bibfnamefont {J.~B.}\ \bibnamefont {Hannon}},
  \bibinfo {author} {\bibfnamefont {R.~M.}\ \bibnamefont {Tromp}}, \bibinfo
  {author} {\bibfnamefont {V.}~\bibnamefont {Perebeinos}}, \bibinfo {author}
  {\bibfnamefont {J.}~\bibnamefont {Tersoff}}, \ and\ \bibinfo {author}
  {\bibfnamefont {F.~M.}\ \bibnamefont {Ross}},\ }\href@noop {} {\bibfield
  {journal} {\bibinfo  {journal} {Nat Mater}\ }\textbf {\bibinfo {volume}
  {11}},\ \bibinfo {pages} {114} (\bibinfo {year} {2012})}\BibitemShut
  {NoStop}%
\bibitem [{\citenamefont {Datta}(1995)}]{DattaBook1995}%
  \BibitemOpen
  \bibfield  {author} {\bibinfo {author} {\bibfnamefont {S.}~\bibnamefont
  {Datta}},\ }\href@noop {} {\emph {\bibinfo {title} {Electronic Transport in
  Mesoscopic Systems}}}\ (\bibinfo  {publisher} {Cambridge University Press,
  Cambridge, England},\ \bibinfo {year} {1995})\BibitemShut {NoStop}%
\bibitem [{\citenamefont {Wiesendanger}(2009)}]{WiesendangerRMP2009}%
  \BibitemOpen
  \bibfield  {author} {\bibinfo {author} {\bibfnamefont {R.}~\bibnamefont
  {Wiesendanger}},\ }\href@noop {} {\bibfield  {journal} {\bibinfo  {journal}
  {Rev. Mod. Phys.}\ }\textbf {\bibinfo {volume} {81}},\ \bibinfo {pages}
  {1495} (\bibinfo {year} {2009})}\BibitemShut {NoStop}%
\bibitem [{\citenamefont {Zhou}\ \emph {et~al.}(2010)\citenamefont {Zhou},
  \citenamefont {Wiebe}, \citenamefont {Lounis}, \citenamefont {Vedmedenko},
  \citenamefont {Meier}, \citenamefont {Blugel}, \citenamefont {Dederichs},\
  and\ \citenamefont {Wiesendanger}}]{ZhouNatPhys2010}%
  \BibitemOpen
  \bibfield  {author} {\bibinfo {author} {\bibfnamefont {L.}~\bibnamefont
  {Zhou}}, \bibinfo {author} {\bibfnamefont {J.}~\bibnamefont {Wiebe}},
  \bibinfo {author} {\bibfnamefont {S.}~\bibnamefont {Lounis}}, \bibinfo
  {author} {\bibfnamefont {E.}~\bibnamefont {Vedmedenko}}, \bibinfo {author}
  {\bibfnamefont {F.}~\bibnamefont {Meier}}, \bibinfo {author} {\bibfnamefont
  {S.}~\bibnamefont {Blugel}}, \bibinfo {author} {\bibfnamefont {P.~H.}\
  \bibnamefont {Dederichs}}, \ and\ \bibinfo {author} {\bibfnamefont
  {R.}~\bibnamefont {Wiesendanger}},\ }\href@noop {} {\bibfield  {journal}
  {\bibinfo  {journal} {Nat. Phys.}\ }\textbf {\bibinfo {volume} {6}},\
  \bibinfo {pages} {187} (\bibinfo {year} {2010})}\BibitemShut {NoStop}%
\bibitem [{\citenamefont {Zhu}\ \emph {et~al.}(2011)\citenamefont {Zhu},
  \citenamefont {Yao}, \citenamefont {Zhang},\ and\ \citenamefont
  {Chang}}]{ZhuPRL2011}%
  \BibitemOpen
  \bibfield  {author} {\bibinfo {author} {\bibfnamefont {J.-J.}\ \bibnamefont
  {Zhu}}, \bibinfo {author} {\bibfnamefont {D.-X.}\ \bibnamefont {Yao}},
  \bibinfo {author} {\bibfnamefont {S.-C.}\ \bibnamefont {Zhang}}, \ and\
  \bibinfo {author} {\bibfnamefont {K.}~\bibnamefont {Chang}},\ }\href@noop {}
  {\bibfield  {journal} {\bibinfo  {journal} {Phys. Rev. Lett.}\ }\textbf
  {\bibinfo {volume} {106}},\ \bibinfo {pages} {097201} (\bibinfo {year}
  {2011})}\BibitemShut {NoStop}%
\bibitem [{\citenamefont {Khajetoorians}\ \emph {et~al.}(2012)\citenamefont
  {Khajetoorians}, \citenamefont {Wiebe}, \citenamefont {Chilian},
  \citenamefont {Lounis}, \citenamefont {Blugel},\ and\ \citenamefont
  {Wiesendanger}}]{KhajetooriansNatPhys2012}%
  \BibitemOpen
  \bibfield  {author} {\bibinfo {author} {\bibfnamefont {A.~A.}\ \bibnamefont
  {Khajetoorians}}, \bibinfo {author} {\bibfnamefont {J.}~\bibnamefont
  {Wiebe}}, \bibinfo {author} {\bibfnamefont {B.}~\bibnamefont {Chilian}},
  \bibinfo {author} {\bibfnamefont {S.}~\bibnamefont {Lounis}}, \bibinfo
  {author} {\bibfnamefont {S.}~\bibnamefont {Blugel}}, \ and\ \bibinfo {author}
  {\bibfnamefont {R.}~\bibnamefont {Wiesendanger}},\ }\href@noop {} {\bibfield
  {journal} {\bibinfo  {journal} {Nat. Phys.}\ }\textbf {\bibinfo {volume}
  {8}},\ \bibinfo {pages} {497} (\bibinfo {year} {2012})}\BibitemShut {NoStop}%
\bibitem [{\citenamefont {Young}\ and\ \citenamefont
  {Kim}(2009)}]{YoungNatPhys2009}%
  \BibitemOpen
  \bibfield  {author} {\bibinfo {author} {\bibfnamefont {A.~F.}\ \bibnamefont
  {Young}}\ and\ \bibinfo {author} {\bibfnamefont {P.}~\bibnamefont {Kim}},\
  }\href@noop {} {\bibfield  {journal} {\bibinfo  {journal} {Nat. Phys.}\
  }\textbf {\bibinfo {volume} {5}},\ \bibinfo {pages} {222} (\bibinfo {year}
  {2009})}\BibitemShut {NoStop}%
\bibitem [{\citenamefont {Williams}\ \emph {et~al.}(2011)\citenamefont
  {Williams}, \citenamefont {Low}, \citenamefont {Lundstrom},\ and\
  \citenamefont {Marcus}}]{WilliamsNatNano2011}%
  \BibitemOpen
  \bibfield  {author} {\bibinfo {author} {\bibfnamefont {J.~R.}\ \bibnamefont
  {Williams}}, \bibinfo {author} {\bibfnamefont {T.}~\bibnamefont {Low}},
  \bibinfo {author} {\bibfnamefont {M.~S.}\ \bibnamefont {Lundstrom}}, \ and\
  \bibinfo {author} {\bibfnamefont {C.~M.}\ \bibnamefont {Marcus}},\
  }\href@noop {} {\bibfield  {journal} {\bibinfo  {journal} {Nat Nano}\
  }\textbf {\bibinfo {volume} {6}},\ \bibinfo {pages} {222} (\bibinfo {year}
  {2011})}\BibitemShut {NoStop}%
\bibitem [{\citenamefont {Rickhaus}\ \emph {et~al.}(2015)\citenamefont
  {Rickhaus}, \citenamefont {Makk}, \citenamefont {Liu}, \citenamefont
  {Tovari}, \citenamefont {Weiss}, \citenamefont {Maurand}, \citenamefont
  {Richter},\ and\ \citenamefont {Schonenberger}}]{RickhausNatComm2015}%
  \BibitemOpen
  \bibfield  {author} {\bibinfo {author} {\bibfnamefont {P.}~\bibnamefont
  {Rickhaus}}, \bibinfo {author} {\bibfnamefont {P.}~\bibnamefont {Makk}},
  \bibinfo {author} {\bibfnamefont {M.-H.}\ \bibnamefont {Liu}}, \bibinfo
  {author} {\bibfnamefont {E.}~\bibnamefont {Tovari}}, \bibinfo {author}
  {\bibfnamefont {M.}~\bibnamefont {Weiss}}, \bibinfo {author} {\bibfnamefont
  {R.}~\bibnamefont {Maurand}}, \bibinfo {author} {\bibfnamefont
  {K.}~\bibnamefont {Richter}}, \ and\ \bibinfo {author} {\bibfnamefont
  {C.}~\bibnamefont {Schonenberger}},\ }\href@noop {} {\bibfield  {journal}
  {\bibinfo  {journal} {Nat. Commun.}\ }\textbf {\bibinfo {volume} {6}},\
  \bibinfo {pages} {7470} (\bibinfo {year} {2015})}\BibitemShut {NoStop}%
\bibitem [{\citenamefont {Ross}\ \emph {et~al.}(2014)\citenamefont {Ross},
  \citenamefont {Klement}, \citenamefont {Jones}, \citenamefont {Ghimire},
  \citenamefont {Yan}, \citenamefont {G.}, \citenamefont {Taniguchi},
  \citenamefont {Watanabe}, \citenamefont {Kitamura}, \citenamefont {Yao},
  \citenamefont {Cobden},\ and\ \citenamefont {Xu}}]{RossNatNano2014}%
  \BibitemOpen
  \bibfield  {author} {\bibinfo {author} {\bibfnamefont {J.~S.}\ \bibnamefont
  {Ross}}, \bibinfo {author} {\bibfnamefont {P.}~\bibnamefont {Klement}},
  \bibinfo {author} {\bibfnamefont {A.~M.}\ \bibnamefont {Jones}}, \bibinfo
  {author} {\bibfnamefont {N.~J.}\ \bibnamefont {Ghimire}}, \bibinfo {author}
  {\bibfnamefont {J.}~\bibnamefont {Yan}}, \bibinfo {author} {\bibfnamefont
  {M.}~\bibnamefont {G.}}, \bibinfo {author} {\bibfnamefont {T.}~\bibnamefont
  {Taniguchi}}, \bibinfo {author} {\bibfnamefont {K.}~\bibnamefont {Watanabe}},
  \bibinfo {author} {\bibfnamefont {K.}~\bibnamefont {Kitamura}}, \bibinfo
  {author} {\bibfnamefont {W.}~\bibnamefont {Yao}}, \bibinfo {author}
  {\bibfnamefont {D.~H.}\ \bibnamefont {Cobden}}, \ and\ \bibinfo {author}
  {\bibfnamefont {X.}~\bibnamefont {Xu}},\ }\href@noop {} {\bibfield  {journal}
  {\bibinfo  {journal} {Nat Nano}\ }\textbf {\bibinfo {volume} {9}},\ \bibinfo
  {pages} {268} (\bibinfo {year} {2014})}\BibitemShut {NoStop}%
\bibitem [{\citenamefont {Baugher}\ \emph {et~al.}(2014)\citenamefont
  {Baugher}, \citenamefont {Churchill}, \citenamefont {Yang},\ and\
  \citenamefont {Jarillo-Herrero}}]{BaugherNatNano2014}%
  \BibitemOpen
  \bibfield  {author} {\bibinfo {author} {\bibfnamefont {B.~W.~H.}\
  \bibnamefont {Baugher}}, \bibinfo {author} {\bibfnamefont {H.~O.~H.}\
  \bibnamefont {Churchill}}, \bibinfo {author} {\bibfnamefont {Y.}~\bibnamefont
  {Yang}}, \ and\ \bibinfo {author} {\bibfnamefont {P.}~\bibnamefont
  {Jarillo-Herrero}},\ }\href@noop {} {\bibfield  {journal} {\bibinfo
  {journal} {Nat Nano}\ }\textbf {\bibinfo {volume} {9}},\ \bibinfo {pages}
  {262} (\bibinfo {year} {2014})}\BibitemShut {NoStop}%
\bibitem [{\citenamefont {Pospischil}\ \emph {et~al.}(2014)\citenamefont
  {Pospischil}, \citenamefont {Furchi},\ and\ \citenamefont
  {Mueller}}]{PospischilNatNano2014}%
  \BibitemOpen
  \bibfield  {author} {\bibinfo {author} {\bibfnamefont {A.}~\bibnamefont
  {Pospischil}}, \bibinfo {author} {\bibfnamefont {M.~M.}\ \bibnamefont
  {Furchi}}, \ and\ \bibinfo {author} {\bibfnamefont {T.}~\bibnamefont
  {Mueller}},\ }\href@noop {} {\bibfield  {journal} {\bibinfo  {journal} {Nat
  Nano}\ }\textbf {\bibinfo {volume} {9}},\ \bibinfo {pages} {257} (\bibinfo
  {year} {2014})}\BibitemShut {NoStop}%
\bibitem [{\citenamefont {Zeng}\ \emph {et~al.}(2013)\citenamefont {Zeng},
  \citenamefont {Morgan}, \citenamefont {Fan}, \citenamefont {Li},
  \citenamefont {Hirono}, \citenamefont {Hu}, \citenamefont {Zhao},
  \citenamefont {Lee}, \citenamefont {Wang}, \citenamefont {Wang},
  \citenamefont {Yu}, \citenamefont {Hawkridge}, \citenamefont {Benamara},\
  and\ \citenamefont {Salamo}}]{ZengAIPAdv2013}%
  \BibitemOpen
  \bibfield  {author} {\bibinfo {author} {\bibfnamefont {Z.}~\bibnamefont
  {Zeng}}, \bibinfo {author} {\bibfnamefont {T.~A.}\ \bibnamefont {Morgan}},
  \bibinfo {author} {\bibfnamefont {D.}~\bibnamefont {Fan}}, \bibinfo {author}
  {\bibfnamefont {C.}~\bibnamefont {Li}}, \bibinfo {author} {\bibfnamefont
  {Y.}~\bibnamefont {Hirono}}, \bibinfo {author} {\bibfnamefont
  {X.}~\bibnamefont {Hu}}, \bibinfo {author} {\bibfnamefont {Y.}~\bibnamefont
  {Zhao}}, \bibinfo {author} {\bibfnamefont {J.~S.}\ \bibnamefont {Lee}},
  \bibinfo {author} {\bibfnamefont {J.}~\bibnamefont {Wang}}, \bibinfo {author}
  {\bibfnamefont {Z.~M.}\ \bibnamefont {Wang}}, \bibinfo {author}
  {\bibfnamefont {S.}~\bibnamefont {Yu}}, \bibinfo {author} {\bibfnamefont
  {M.~E.}\ \bibnamefont {Hawkridge}}, \bibinfo {author} {\bibfnamefont
  {M.}~\bibnamefont {Benamara}}, \ and\ \bibinfo {author} {\bibfnamefont
  {G.~J.}\ \bibnamefont {Salamo}},\ }\href@noop {} {\bibfield  {journal}
  {\bibinfo  {journal} {AIP Advances}\ }\textbf {\bibinfo {volume} {3}},\
  \bibinfo {eid} {072112} (\bibinfo {year} {2013})}\BibitemShut {NoStop}%
\bibitem [{\citenamefont {Bathon}\ \emph {et~al.}(2015)\citenamefont {Bathon},
  \citenamefont {S.Achilli}, \citenamefont {P.Sessi}, \citenamefont
  {V.A.Golyashov}, \citenamefont {K.A.Kokh}, \citenamefont {O.E.Tereshchenko},\
  and\ \citenamefont {M.Bode}}]{BathonArxiv2015}%
  \BibitemOpen
  \bibfield  {author} {\bibinfo {author} {\bibfnamefont {T.}~\bibnamefont
  {Bathon}}, \bibinfo {author} {\bibnamefont {S.Achilli}}, \bibinfo {author}
  {\bibnamefont {P.Sessi}}, \bibinfo {author} {\bibnamefont {V.A.Golyashov}},
  \bibinfo {author} {\bibnamefont {K.A.Kokh}}, \bibinfo {author} {\bibnamefont
  {O.E.Tereshchenko}}, \ and\ \bibinfo {author} {\bibnamefont {M.Bode}},\
  }\href@noop {} {\bibfield  {journal} {\bibinfo  {journal} {arXiv:1512.06554
  [cond-mat.mes-hall]}\ } (\bibinfo {year} {2015})}\BibitemShut {NoStop}%
\bibitem [{\citenamefont {Maze}\ \emph {et~al.}(2008)\citenamefont {Maze},
  \citenamefont {Stanwix}, \citenamefont {Hodges}, \citenamefont {Hong},
  \citenamefont {Taylor}, \citenamefont {Cappellaro}, \citenamefont {Jiang},
  \citenamefont {Dutt}, \citenamefont {Togan}, \citenamefont {Zibrov},
  \citenamefont {Yacoby}, \citenamefont {Walsworth},\ and\ \citenamefont
  {Lukin}}]{MazeNature2008}%
  \BibitemOpen
  \bibfield  {author} {\bibinfo {author} {\bibfnamefont {J.~R.}\ \bibnamefont
  {Maze}}, \bibinfo {author} {\bibfnamefont {P.~L.}\ \bibnamefont {Stanwix}},
  \bibinfo {author} {\bibfnamefont {J.~S.}\ \bibnamefont {Hodges}}, \bibinfo
  {author} {\bibfnamefont {S.}~\bibnamefont {Hong}}, \bibinfo {author}
  {\bibfnamefont {J.~M.}\ \bibnamefont {Taylor}}, \bibinfo {author}
  {\bibfnamefont {P.}~\bibnamefont {Cappellaro}}, \bibinfo {author}
  {\bibfnamefont {L.}~\bibnamefont {Jiang}}, \bibinfo {author} {\bibfnamefont
  {M.~V.~G.}\ \bibnamefont {Dutt}}, \bibinfo {author} {\bibfnamefont
  {E.}~\bibnamefont {Togan}}, \bibinfo {author} {\bibfnamefont {A.~S.}\
  \bibnamefont {Zibrov}}, \bibinfo {author} {\bibfnamefont {A.}~\bibnamefont
  {Yacoby}}, \bibinfo {author} {\bibfnamefont {R.~L.}\ \bibnamefont
  {Walsworth}}, \ and\ \bibinfo {author} {\bibfnamefont {M.~D.}\ \bibnamefont
  {Lukin}},\ }\href@noop {} {\bibfield  {journal} {\bibinfo  {journal}
  {Nature}\ }\textbf {\bibinfo {volume} {455}},\ \bibinfo {pages} {644}
  (\bibinfo {year} {2008})}\BibitemShut {NoStop}%
\bibitem [{\citenamefont {Taylor}\ \emph {et~al.}(2008)\citenamefont {Taylor},
  \citenamefont {Cappellaro}, \citenamefont {Childress}, \citenamefont {Jiang},
  \citenamefont {Budker}, \citenamefont {Hemmer}, \citenamefont {Yacoby},
  \citenamefont {Walsworth},\ and\ \citenamefont {Lukin}}]{TaylorNatPhys2008}%
  \BibitemOpen
  \bibfield  {author} {\bibinfo {author} {\bibfnamefont {J.~M.}\ \bibnamefont
  {Taylor}}, \bibinfo {author} {\bibfnamefont {P.}~\bibnamefont {Cappellaro}},
  \bibinfo {author} {\bibfnamefont {L.}~\bibnamefont {Childress}}, \bibinfo
  {author} {\bibfnamefont {L.}~\bibnamefont {Jiang}}, \bibinfo {author}
  {\bibfnamefont {D.}~\bibnamefont {Budker}}, \bibinfo {author} {\bibfnamefont
  {P.~R.}\ \bibnamefont {Hemmer}}, \bibinfo {author} {\bibfnamefont
  {A.}~\bibnamefont {Yacoby}}, \bibinfo {author} {\bibfnamefont
  {R.}~\bibnamefont {Walsworth}}, \ and\ \bibinfo {author} {\bibfnamefont
  {M.~D.}\ \bibnamefont {Lukin}},\ }\href@noop {} {\bibfield  {journal}
  {\bibinfo  {journal} {Nat. Phys.}\ }\textbf {\bibinfo {volume} {4}},\
  \bibinfo {pages} {810} (\bibinfo {year} {2008})}\BibitemShut {NoStop}%
\bibitem [{\citenamefont {Zhao}\ \emph {et~al.}(2011)\citenamefont {Zhao},
  \citenamefont {Hu}, \citenamefont {Ho}, \citenamefont {Wan},\ and\
  \citenamefont {Liu}}]{ZhaoNatNano2011}%
  \BibitemOpen
  \bibfield  {author} {\bibinfo {author} {\bibfnamefont {N.}~\bibnamefont
  {Zhao}}, \bibinfo {author} {\bibfnamefont {J.-L.}\ \bibnamefont {Hu}},
  \bibinfo {author} {\bibfnamefont {S.-W.}\ \bibnamefont {Ho}}, \bibinfo
  {author} {\bibfnamefont {J.~T.~K.}\ \bibnamefont {Wan}}, \ and\ \bibinfo
  {author} {\bibfnamefont {R.-B.}\ \bibnamefont {Liu}},\ }\href@noop {}
  {\bibfield  {journal} {\bibinfo  {journal} {Nat. Nanotechnol.}\ }\textbf
  {\bibinfo {volume} {6}},\ \bibinfo {pages} {242} (\bibinfo {year}
  {2011})}\BibitemShut {NoStop}%
\bibitem [{\citenamefont {Shi}\ \emph {et~al.}(2014)\citenamefont {Shi},
  \citenamefont {Kong}, \citenamefont {Wang}, \citenamefont {Kong},
  \citenamefont {Zhao}, \citenamefont {Liu},\ and\ \citenamefont
  {Du}}]{ShiNatPhys2014}%
  \BibitemOpen
  \bibfield  {author} {\bibinfo {author} {\bibfnamefont {F.}~\bibnamefont
  {Shi}}, \bibinfo {author} {\bibfnamefont {X.}~\bibnamefont {Kong}}, \bibinfo
  {author} {\bibfnamefont {P.}~\bibnamefont {Wang}}, \bibinfo {author}
  {\bibfnamefont {F.}~\bibnamefont {Kong}}, \bibinfo {author} {\bibfnamefont
  {N.}~\bibnamefont {Zhao}}, \bibinfo {author} {\bibfnamefont {R.-B.}\
  \bibnamefont {Liu}}, \ and\ \bibinfo {author} {\bibfnamefont
  {J.}~\bibnamefont {Du}},\ }\href@noop {} {\bibfield  {journal} {\bibinfo
  {journal} {Nat. Phys.}\ }\textbf {\bibinfo {volume} {10}},\ \bibinfo {pages}
  {21} (\bibinfo {year} {2014})}\BibitemShut {NoStop}%
\end{thebibliography}

%

\end{document}